\documentstyle[epsf,sprocl]{article}

\input epsf
\input{psfig.sty}
\newcommand{\beqn} {\begin{equation}}
\newcommand{\eqn} {\end{equation}}

\def\PRL{\em Phys. Rev. Lett.}

\def\NP{{Nucl.\ Phys.\ }}

\def\PL{{Phys.\ Lett.\ }}
\def\PR{{Phys.\ Rev.\ }}
\def\PRL{{Phys.\ Rev.\ Lett.\ }}

\def\be{\begin{equation}}
\def\ee{\end{equation}}
\def\bea{\begin{eqnarray}}
\def\eea{\end{eqnarray}}

\begin{document}

\title{SCREENING LENGTHS IN SU(2) GAUGE THEORY AT FINITE 
TEMPERATURE\footnote{Contribution to the proceedings of the conference on
STRONG AND ELECTROWEAK MATTER '97, 21-25 May 1997, Eger, Hungary} }

\author{
\vskip -130pt
 \mbox{} \hfill BI-TP 97/30\\
 \mbox{} \hfill August 1997\\
\vskip 100pt
U. M. HELLER$^a$, F. KARSCH$^b$ and J. RANK$^{a,b}$}

\address{~\\
$^a$ SCRI, Florida State University, Tallahassee, FL 32306-4130, USA\\
$^b$ Fakult\"at f\"ur Physik, Universit\"at Bielefeld,
D-33615, Bielefeld, Germany}

\vspace*{-1.0cm}
\maketitle\abstracts{
We discuss the calculation of electric and magnetic screening
masses in SU(2) gauge theory. The temperature dependence of these
masses obtained from the long-distance behaviour of spatial correlation functions 
has been analyzed for temperatures up to $10^4 T_c$.}

\section{Introduction}
One of the basic concepts that guide our intuition about the properties 
of the high temperature plasma phase of QCD is the occurrence of 
chromo-electric and -magnetic screening. 
From the early perturbative calculations at high temperatures \cite{KapB}
we know that a non-vanishing electric screening mass, $m_e$, is needed to
control the infrared behaviour of QCD at momentum scales of ${\cal O} (gT)$. 
Although mechanisms have been suggested which do not require the dynamic  
generation of a magnetic mass scale, $m_m \sim {\cal O} (g^2 T)$, which 
could cure the    
remaining infrared divergences \cite{BlIa96} the existence of such a 
mass would clearly be sufficient \cite{Lin80}. 
Assuming the existence of a non-vanishing
magnetic mass Rebhan has shown that this will 
influence the perturbative calculation of the electric mass already at 
next-to-leading order \cite{Re93}, {\it i.e.} at ${\cal O} (g^2\ln g)$. 

The analysis of electric and magnetic screening properties in the high
temperature phase also is important for our understanding of the nature of 
fundamental excitations in the QCD plasma phase. Are quarks and gluons
the basic degrees of freedom in the plasma phase? Can one give to them
a gauge invariant meaning or should one try to understand the plasma
phase in terms of colourless excitations only? Calculations
of the QCD equation of state clearly suggests that the relevant degrees
of freedom in the high temperature phase are those of quarks and gluons.
However, to which extent these partonic degrees of freedom do have
further dynamic significance in the high temperature phase is not obvious.
Can we give, for instance, the thermal electric gluon mass a physical, i.e.
gauge invariant, meaning? To some extent this has been answered by Kobes et.
al. \cite{KoKuRe90}. They show that although the gluon propagator,
\begin{equation}
G_\mu (p_0,\vec{p}) \equiv \langle {\rm Tr} A_\mu (p_0,\vec{p}) 
A_\mu^{\dagger} (p_0,\vec{p}) \rangle 
\sim \bigl(p^2 + \Pi_{\mu \mu} (p_0,\vec{p}) \bigr)^{-1} \quad~,
\end{equation}
is a gauge dependent observable, the pole masses, 
\begin{equation}
m_\mu^2 = \Pi_{\mu \mu} (0, |\vec{p}|^2=-m_\mu^2) \quad ~
\label{polemass}
\end{equation}
are gauge invariant.  

In order to circumvent the definition of $m_e$ and $m_m$ in terms of gauge 
dependent operators attempts have been undertaken to introduce
gauge invariant observables for the calculation of 
gluon screening masses \cite{ArYa95}. Here, however, one has to examine 
in how far the masses extracted from gauge invariant operators correspond 
to those of {\it elementary excitations} or to {\it quasi-particle states} which 
may result from superpositions of several gluons. This problem became, 
for instance, apparent in recent studies of the thermal W-boson mass in the 
symmetric high temperature phase of the electroweak theory \cite{KaNePaRa95}.

Eventually we clearly have to aim at an analysis of various operators that
allow to extract the thermal gluon masses.  In the following we will 
concentrate on the  
calculation of electric, $m_e\equiv m_0$, and magnetic, $m_m \equiv m_i~,~i\ne 0$, 
masses from the gluon propagator in Landau gauge. These pole masses can 
be obtained from the exponential decay of finite temperature gluon correlation 
functions at large spatial separations. 

\section{Electric and Magnetic Screening Masses}

\begin{figure}[htb]
\begin{center}
\hskip 0.5truecm\epsffile{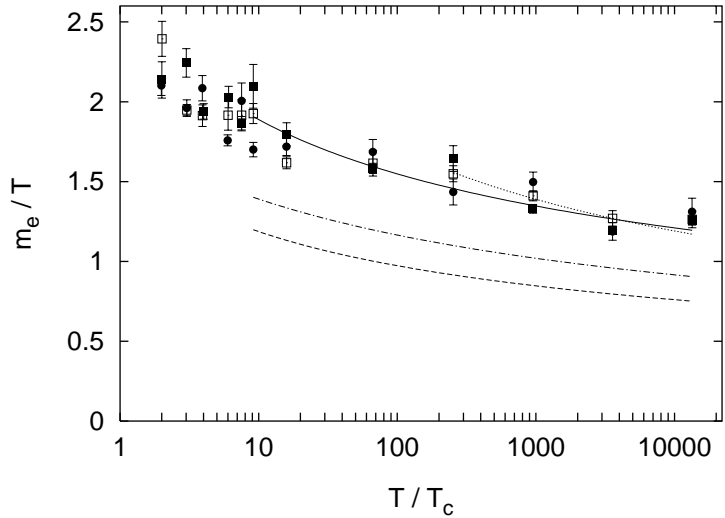}
\end{center}
\caption{The electric screening mass in units of the temperature 
versus $T/T_c$ calculated on lattices
of size $4\times 32^3$ (filled squares) and $8 \times 32^2\times 64$ (open
squares) with the Wilson action and on a $4\times 32^3$ lattice with the 
Symanzik improved action (filled circles). The curves show leading order
(dashed) and next-to-leading order (dashed-dotted) perturbative results as
well as fits with the ans\"atze given in Eq.~(4) (see text for details). 
\label{fig:memm}}
\end{figure}

We have analyzed the gluon propagator in coordinate space, {\it i.e.} we 
calculate spatial correlation functions of static ($p_0\equiv 0$) gauge fields, 
$\tilde{A}_\mu (x_3) \equiv \sum_{x_0,x_1,x_3} A_\mu (x_0,x_1,x_2,x_3)$,
in the $x_3$-direction of finite temperature lattices of size $N_\tau
\times N_\sigma^2\times  N_3$. The long distance behaviour of the correlation 
function,
$\tilde{G}_\mu (x_3) = \langle {\rm Tr} \tilde{A}_\mu (x_3) 
\tilde{A}_\mu^{\dagger} (x_3) \rangle$, 
yields the electric and magnetic masses, respectively~\footnote{These 
calculations are an extension of our earlier
investigations in a much smaller temperature interval~\cite{HeKaRa95}.
For details
on the definition of  $A_\mu (x)$ on the lattice, the gauge fixing and
further lattice specific details we refer to Refs. 8  and 9.}.
In order to check the influence of discretization errors
resulting from the finite lattice spacing, $a$, we have performed
calculations with the standard Wilson action as well as with an ${\cal O} (a^2)$
Symanzik-improved action. In addition we have chosen two different temporal 
lattice sizes, 
$N_\tau=4$ and 8. At fixed temperature $T\equiv
1/N_\tau a$ we thus can perform calculations at values of the lattice cut-off that
differ by a factor two. 

Let us start with a discussion of the electric screening mass.
In Fig.~\ref{fig:memm} we show $m_e/T$ for both types of actions and the 
two different temporal lattice sizes. Within errors,
$m_e/T$ does not differ significantly for the three cases. 
Even the Symanzik-improved action, 
does not shift the electric screening mass in any direction.
This is quite different from what has been observed in calculations of 
bulk thermodynamic observables like the energy density or pressure~\cite{BeKaLa96}. It is, however, in accordance with the expectation that the
screening masses are entirely dominated by infra-red effects, while the
bulk thermodynamic observables receive large contributions from 
ultra-violet modes, which in turn are strongly influenced by finite cut-off
effects.

We have analyzed the temperature dependence of $m_e/T$ using ans\"atze
motivated by the leading and next-to-leading order perturbative 
calculations
\begin{equation}
\biggl({m_e \over T} \biggr)^2 = \cases{Ag^2(T) &,~case-A \cr
{2\over 3} g^2(T) \biggl[ 1 + {3 \over
2\pi} {m_e\over T} \biggl(\ln\bigl({2m_e\over m_m}\bigr) - {1\over 2} \biggr) 
 \biggr] + B g^4(T) &,~case-B\cr
} 
\label{fits}
\end{equation}
where we use for the running coupling the two-loop $\beta$-function 
\begin{equation}
g^{-2} (T) = {11 \over 12\pi^2} \ln \biggl({2\pi T\over
\Lambda_{\overline{\rm MS}}}\biggr) + {17 \over 44 \pi^2} \ln\biggl[ 2\ln
\biggl( {2\pi T\over \Lambda_{\overline{\rm MS}}}\biggr) \biggr] \quad ,
\end{equation}
and relate $\Lambda_{\overline{\rm MS}}$ to the critical temperature
for the deconfinement transition, 
$T_c/\Lambda_{\overline{\rm MS}} \simeq 1.08$~\cite{FiHeKa93}.

\begin{figure}[htb]
\hskip 1.3truecm\epsffile{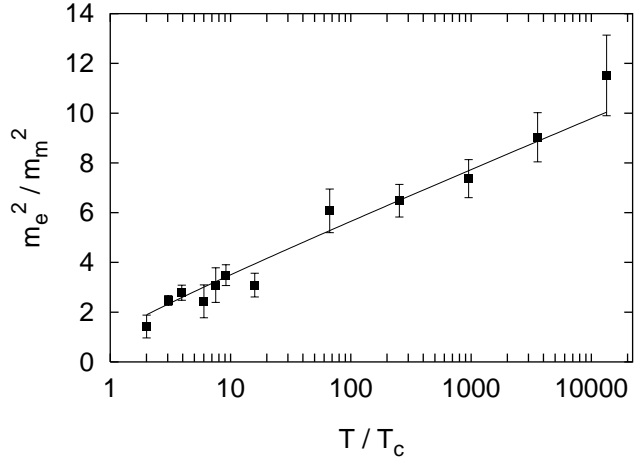}
\caption{Squared ratio of the electric and magnetic screening masses vs.\ $T/T_c$
from simulations on a $8\times 32^2 \times 64$ lattice using the Wilson
action.}
\label{fig:m64x8}
\end{figure}

To make use of the next-to-leading order ansatz we determine also the
magnetic mass, $m_m$. Results for the ratio $m_e/m_m$ obtained from
calculations with the Wilson action on a lattice of size
$8\times 32^2 \times 64$ are shown in Fig.~\ref{fig:m64x8}. The naive
expectation, $m_m/m_e \sim g (T)$, does seem to describe this ratio quite
well. A fit with such an ansatz for $T\ge 2 T_c$ yields
\beqn
\left( \frac{m_e}{m_m} \right)^2 = (7.4 \pm 0.3)\;
g^{-2}(T)
\quad .
\label{memm_ratio}
\eqn
with $\chi^2 / \mbox{dof} = 1.4$.
A fit for $m_m$ itself, using the ansatz $m_m/T \sim g^2(T)$,
yields
\beqn
\frac{m_m}{T} = (0.46 \pm 0.01)g^2 (T) \quad ,
\label{meT_ratio}
\eqn
which is in good agreement with our earlier calculation in a much narrower
temperature interval \cite{HeKaRa95}.

In Fig.~\ref{fig:memm} we show the leading (case-A with A=2/3)
and next-to-leading order (case-B with B=0) perturbative results as dashed
and dash-dotted curves, respectively. Even at temperatures $T \sim 10^4 T_c$ 
the leading order result deviates from the numerical results by nearly a factor 
of two. The next-to-leading order result clearly
yields an improved description. Including an additional ${\cal O}(g^4T)$ 
correction as suggested in the case-B fit leads, however, to a too
strong variation with temperature to provide a good description in the 
entire temperature interval. This is shown by the dotted curve in
Fig.~\ref{fig:memm}. The fit for temperatures $T > 100 T_c$ yields for the 
coefficient of the 
${\cal O} (g^4)$ correction $B= 0.54\pm 0.03$. The case-A fit, 
on the other hand, does yield a satisfactory description of the data 
for $T > 10T_c$. This is 
also shown in Fig.~\ref{fig:memm} as a solid curve. For the fit parameter we find
$A=1.69\pm 0.02$ with $\chi^2 / \mbox{dof} = 4.1$. This, of course, is 
consistent with the fits from Eq.~\ref{memm_ratio} and \ref{meT_ratio}. 
The numerical value, however, exceeds the leading order perturbative result
by a factor 2.5.      

\section{Conclusions}
\vspace*{-0.1cm}
The analysis of electric and magnetic screening masses shows that
$m_e/T$ as well as $m_m/T$ are {\it running with temperature}. The 
temperature variation is consistent with a logarithmic dependence. In
particular, we have evidence that $m_m/m_e \sim g(T)$ as expected from  
general considerations of the infrared-behaviour of high
temperature QCD. Quantitative results, however, do not agree with
leading 
and next-to-leading order perturbation theory even at temperatures as
high as $10^4 T_c$. Similar conclusions have been drawn from
investigations of the SU(2) finite temperature theory in the context
of dimensional reduction using a gauge invariant operator \cite{KaLaRuSh97,Ru97}.

\end{document}